\documentclass[11pt]{article}
\usepackage[dvipsnames]{xcolor} 
\usepackage[hidelinks]{hyperref}
\usepackage{enumitem}
\usepackage{amsmath,amssymb,epsf,cite,graphicx,subfigure}
\usepackage{amsmath,braket,tikzsymbols}
\usepackage[bbgreekl]{mathbbol}
\usepackage{bbold}

\numberwithin{equation}{section}

\definecolor{airforceblue}{rgb}{0.36, 0.54, 0.66}
\newcommand{\beq}{\begin{equation}}
\newcommand{\eeq}{\end{equation}}

\usepackage[mathscr]{euscript}

\begin{document}
\baselineskip=15.5pt
\pagestyle{plain}
\setcounter{page}{1}

\begin{center}
{\LARGE \bf Eternal traversable wormholes in three dimensions}
\vskip 1cm

\textbf{William Harvey$^{a}$ and Kristan Jensen$^{b}$}

\vspace{0.5cm}

{\small Department of Physics and Astronomy, University of Victoria, Victoria, BC V8W 3P6, Canada\\}

\vspace{0.3cm}

{\tt \small ${}^a$wharvey@uvic.ca,}
{\tt  \small ${}^b$kristanj@uvic.ca \\}

\medskip

\end{center}

\vskip1cm

\begin{center}
{\bf Abstract}
\end{center}

We consider three-dimensional gravity with negative cosmological constant coupled to a large number of light matter fields dual to relevant operators. By imposing suitable boundary conditions on the matter fields we find eternal traversable wormhole deformations of the BTZ black hole, leading to a three-dimensional analogue of the AdS$_2$ eternal traversable wormhole found by Maldacena and Qi. We further identify the field theory of boundary gravitons in this setting, which we then use to compute the spectrum of gravitational fluctuations.

\hspace{.3cm}

\newpage

\tableofcontents

\section{Introduction}

The subject of spacetime wormholes in gravity has been long, rich, and confusing (see e.g.~\cite{Coleman:1988tj}). In Lorentzian signature, maximally extended black hole geometries are examples of non-traversable wormholes, where the space behind the horizon smoothly connects the two exteriors of the black hole. Traversable wormholes, however, are abhorred by gravity. Traversable wormhole solutions to the Einstein's equations are forbidden by the averaged null energy condition~\cite{Morris:1988tu}, and in ad hoc models of gravity coupled to exotic matter designed to support a traversable wormhole, the ensuing solution is usually afflicted by pathologies. In Euclidean signature, wormholes are in general hard to come by. There are wormhole solutions in models of gravity coupled to $p$-form gauge fields~\cite{Maldacena:2004rf,Marolf:2021kjc}, including the axion wormholes of Giddings and Strominger~\cite{Giddings:1989bq}. However these geometries do not lead to traversable wormholes upon continuation to Lorentzian signature; are difficult to embed into string theory; and, to this day, it is not known if the successful embeddings are stable. 

The status of wormholes in quantum gravity, that is in an off-shell formulation of gravity, is a matter of ongoing research. In non-perturbatively tractable models of quantum gravity like worldsheet string theory and Jackiw-Teitelboim gravity (itself likely a corner in the landscape of worldsheet string theories~\cite{Saad:2019lba}) one sums over all metrics consistent with the boundary conditions. A sum over metrics in higher spacetime dimension would include a sum over Euclidean wormholes in Euclidean signature and traversable wormholes in Lorentzian signature. The physics of these off-shell configurations is yet unclear. As they are not solutions to the field equations, a wormhole is really one point in a sum over wormhole metrics, and this sum is dominated by those geometries of least action, where the wormhole pinches off and the geometry becomes strongly curved. That is, the sum over these off-shell configurations is sensitive to the ultraviolet completion of quantum gravity and, without help, inaccessible in gravitational effective field theory~\cite{Cotler:2021cqa,Cotler:2022rud}.

The role of wormholes takes on new life in the AdS/CFT correspondence. Two-sided black holes are dual to thermofield double states~\cite{} of the dual conformal field theory (CFT), where the non-traversable Einstein-Rosen bridge between the two exteriors encodes entanglement between the left and right copies of the CFT. The absence of traversable wormhole solutions connecting two asymptotic boundaries dovetails with the fact that, if such a traversable wormhole existed, it would violate boundary causality. In Euclidean signature the paucity (although not complete absence~\cite{Maldacena:2004rf,Marolf:2021kjc,Loges:2023ypl}) of wormhole solutions is consistent with boundary factorization, and as far as off-shell (and certain singular~\cite{Saad:2018bqo}) configurations go, there is evidence that these encode the level statistics of heavy CFT microstates dual to black holes~\cite{Saad:2018bqo,Cotler:2021cqa}.

As mentioned above, the absence of traversable wormhole solutions in AdS/CFT can be understood from the boundary as a consequence of causality, the statement that operators on different boundaries commute with each other. In the bulk it follows from the averaged null energy condition (ANEC), which is obeyed for perturbative AdS vacua in the string landscape. However, phrasing the matter this way suggests a way of making traversable wormholes, as recognized by Gao, Jafferis, and Wall~\cite{Gao:2016bin}. From the bulk point of view we need a controlled violation of the ANEC, but from the boundary point of view we need to couple different asymptotic boundaries to each other. 

The simplest setting in which one can realize this idea is with two asymptotic boundaries and where one considers a phenomenological model of gravity coupled to a very large number $N_f = O(1/G)$ (with $G$ the gravitational constant) of light matter fields $\phi^i$ dual to relevant operators $O_L^i$ and $O_R^i$ on the two boundaries. If the bulk masses are sufficiently low that the double trace $O^2$ is relevant, we then add a negative non-local double trace deformation $-\sum_{i=1}^{N_f}\int dt d^dx \,g(t,x) O^i_L(t,x)O^i_R(t,x)$ which directly couples the two boundaries. In the bulk this corresponds to modified boundary conditions on the $\phi^i$, which leads to an $O(1)$ violation of the ANEC for each field, and so an effective bulk stress tensor of $O(N_f)=O(1/G)$. This is comparable to the curvature contribution to the Einstein's equations and so can support a macroscopic wormhole. Models of this sort should be regarded as purely phenomenological, as they have yet to be embedded into string theory, and the ensuing wormholes are of conceptual interest.

In their original work, Gao, Jafferis, and Wall studied such models when the double trace deformation is turned on for a short interval of time, opening up causal connection (although not perfect transmission of signals) between two asymptotic boundaries for a window of time. Shortly thereafter Maldacena and Qi~\cite{Maldacena:2018lmt} studied an eternal version of the Gao-Jafferis-Wall deformation in JT gravity, with constant double trace coupling $g$, matching it against the low-energy limit of two Sachdev-Ye-Kitaev models coupled by a similar ``double trace'' interaction between the microscopic fermions of the two copies.

In this paper we construct and study eternal traversable wormholes in three spacetime dimensions connecting two asymptotically AdS boundaries. These wormholes are supported by an eternal negative double trace deformation coupling the two boundaries, a higher-dimensional version of the wormholes of Maldacena and Qi. 

Our construction relies heavily on the previous work of one of us with Cotler~\cite{Cotler:2020ugk}, in which pure three-dimensional gravity with negative cosmological constant on certain spacetimes can be written entirely as a boundary model. For Einstein gravity on spacetimes with the topology of global AdS$_3$~\cite{Cotler:2018zff}, the resulting boundary model is a higher-dimensional version of the Schwarzian theory~\cite{Jensen:2016pah,Maldacena:2016upp,Engelsoy:2016xyb} that governs the boundary dynamics of JT gravity. For Euclidean wormholes with the topology of a torus times an interval (i.e. where the boundary is two disconnected tori) one finds a more complicated boundary model, from which one can compute the wormhole amplitude to one-loop order. This latter model is the one relevant for us. Before adding matter fields, this amplitude lacks a saddle-point approximation. Decompactifying the Euclidean time direction and analytically continuing it to Lorentzian signature, so as to describe Lorentzian traversable wormholes, we then deform this model so as to account for a large number of matter fields and double trace deformation. Our approach, similar to that taken by Maldacena and Qi, is a little indirect. Rather than computing the effective matter stress tensor directly, we use that, to leading order at small $G$, we can replace both by an insertion of $\exp\left( \sum_{i=1}^{N_f} \int dt dx \,g \,\langle O_L^i(t,x)O_R^i(t,x)\rangle\right)$ into the sum over metrics, with $\langle O_L^i O_R^i\rangle$ the boundary-to-boundary left-right two-point function of $O$ in the wormhole. The term in the exponent deforms the wormhole action in such a way as to now admit saddle points, which from the bulk point of view are three-dimensional traversable wormholes supported by an eternal Gao-Jafferis-Wall deformation.

The main virtue of our approach, besides allowing us to find these wormholes, is that we can probe some of their physics directly from this boundary action. For example, we study the spectrum of gravitational fluctuations, which allows us to assess the perturbative stability of the wormholes we find. In particular we find that gravitational perturbations are stable as long as the double trace deformation is relevant. In principle we may also use this model to compute the gravitational contribution to scattering in the wormhole geometry. 

These wormholes are also related to the Euclidean wormholes discussed in~\cite{Chandra:2022bqq}. One example those authors considered is a two-torus wormhole where a particle worldline connects the two boundaries and supports the wormhole. The particle is putatively dual to a heavy  single-trace operator below the BTZ threshold with dimension $\Delta = O(c)$, and the worldline to a Gao-Jafferis-Wall deformation, which is evidently enough to hold the wormhole open. Our configuration effectively has a smeared version of the deformation in the regime with $\Delta = O(1)$.

The remainder of this manuscript is organized as follows. In Section~\ref{S:review} we gather some useful facts about three-dimensional gravity and Euclidean wormholes therein that allow us to construct the boundary action for our real-time traversable wormholes. We find wormhole solutions of the latter in Section~\ref{S:wormholes} as well as study the spectrum of gravitational fluctuations. In Section~\ref{S:3d} we study the graviton propagator for these wormholes, which in principle allows for a computation of bulk scattering, and a plausible reconstruction of the three-dimensional wormhole spacetime when the fields $\phi^i$ are scalars. We wrap up with a Discussion in Section~\ref{S:discuss}.

\section{Preliminaries}
\label{S:review}

\subsection{First-order formulation of three-dimensional gravity}

Pure three-dimensional gravity with negative cosmological constant is described by the action
\beq
\label{3DEinsteinHilbert}
	S = \frac{1}{16 \pi G} \int d^{3}x \sqrt{-g} \left(R + \frac{2}{\ell^{2}} \right) + (\text{bdy})\,,
\eeq
for an appropriate boundary term. We henceforth choose $\ell=1$ units. The basic solution to the field equations of the model is global AdS$_3$,
\beq
\label{globalAdS3}
    ds^{2} = -\cosh^{2}(\rho) dt^{2} + \sinh^{2}(\rho) dx^{2} + d \rho^{2}\,,
\eeq
with $x\sim x+2\pi$. 

In the first-order formulation of gravity we decompose the metric into a dreibein $e^A_{\mu}$ via
\beq
	g_{\mu\nu} = \eta_{AB} e^A_{\mu}e^B_{\nu}\,,
\eeq
with $A,B=0,1,2$. This decomposition introduces a redundancy under local Lorentz rotations $e^A_{\mu} \to \Lambda^A{}_B e^B_{\mu}$. We then divide by that redundancy and introduce an associated spin connection $\omega^A{}_{B\mu}$, obeying $\omega_{(AB)\mu} = 0$ (flat indices $A$ and $B$ are raised and lowered with the Minkowski metric $\eta_{AB}$). The ensuing action is
\beq
	\label{firstordergrav}
    S = -\frac{1}{16 \pi G} \int \epsilon_{ABC} \, e^{A} \wedge \bigg(d \omega^{BC} + \omega^{B}{}_D \wedge \omega^{DC} + \frac{1}{3} e^{B} \wedge e^{C} \bigg) + (\text{bdy})\,,
\eeq
where $e^A = e^A_{\mu}dx^{\mu}$ and $\omega^A{}_B = \omega^A{}_{B\mu}dx^{\mu}$. The spin connection appears quadratically and so can be integrated out. Doing so enforces that the spacetime is torsion-free, and plugging the result back into the action one recovers Einstein gravity in the second-order formalism. Note that this action only has a single time derivative, which implies it is in Hamiltonian form. 

It is useful~\cite{Achucarro:1986uwr,Witten:1988hc} to combine the dreibein and spin connection into the variables
\beq
	A^A = \frac{1}{2}\epsilon^{ABC}\omega_{BC} + e^A \,, \qquad \bar{A}^A = \frac{1}{2}\epsilon^{ABC}\omega_{BC}-e^A\,,
\eeq
and then letting $J_A$ and $\bar{J}_A$ be the generators of two copies of $\mathfrak{sl}(2;\mathbb{R})$ in the fundamental representation, satisfying
\beq
	[J_A,J_B]=\epsilon_{ABC} J^C\,, \qquad \text{tr}\left( J_A J_B\right) = \frac{1}{2}\eta_{AB}\,,
\eeq
and similarly for the $\bar{J}_A$'s, to define matrix-valued one-forms $A = A^AJ_A$ and $\bar{A} = \bar{A}^A\bar{J}_A$. In terms of these variables the first-order action becomes
\beq
	S = - \frac{k}{4\pi} \int \left( I[A]-I[\bar{A}]\right)+(\text{bdy})\,, \quad k = \frac{1}{4G}\,, \quad I[A] = \text{tr}\left( A \wedge dA + \frac{2}{3}A\wedge A \wedge A\right)\,,
\eeq
resembling a Chern-Simons theory with algebra $\mathfrak{sl}(2;\mathbb{R})\times \mathfrak{sl}(2;\mathbb{R})$. Indeed the equations of motion that come from varying this action with respect to $A$ and $\bar{A}$,
\beq
	F= dA + A \wedge A = 0\,, \qquad \bar{F} = d\bar{A} + \bar{A}\wedge \bar{A} = 0\,,
\eeq 
are precisely the torsion-free and Einstein's equations expressed in terms of $A$ and $\bar{A}$. In fact this resemblance goes beyond the action and equations of motion: on a solution to the field equations, linearized diffeomorphisms and local Lorentz transformations act on $A$ and $\bar{A}$ in the same way as infinitesimal $\mathfrak{sl}(2;\mathbb{R})\times \mathfrak{sl}(2;\mathbb{R})$ gauge transformations. These results can be rephrased as the statement that 3d gravity is classically equivalent to Chern-Simons theory. One can do a bit better and show a quantum mechanical equivalence between gravity on a solid cylinder, i.e. the sum over metrics continuously connected to the global AdS$_3$ saddle, is equivalent to a winding sector of an $SO(2,2)$ Chern-Simons theory, where $SO(2,2)$ is a particular global completion of $\mathfrak{sl}(2;\mathbb{R})\times\mathfrak{sl}(2;\mathbb{R})$. See~\cite{Castro:2011iw,Cotler:2020ugk} for details. However 3d gravity and Chern-Simons theory differ non-perturbatively; for example the torus times interval amplitude of 3d gravity computed in~\cite{Cotler:2020ugk} is finite, whereas the corresponding amplitude diverges in $SO(2,2)$ Chern-Simons theory.

\subsection{Boundary action}

In this work we are interested in traversable wormholes with the topology of an annulus times time. These wormholes are closely related to the Euclidean wormholes considered in~\cite{Cotler:2020ugk}, which have the topology of an annulus times (compact) Euclidean time. In that work gravity on such a space was reduced to a boundary theory. (See also~\cite{Eberhardt:2022wlc} for related work.)

In the present work we analytically continue that theory back to Lorentzian signature and add an eternal double trace deformation. The ensuing action, which we describe in Section~\ref{S:wormholeAction}, has saddles that correspond to spacetime wormholes. To arrive at it we take an excursion and review the derivation in~\cite{Cotler:2020ugk} of the boundary action for Euclidean wormholes of the form annulus times time. (See also~\cite{Henneaux:2019sjx} for a related derivation in the context of the two-sided BTZ black hole.) The main result is Eq.~\eqref{EuclideanWormhole}, the action for a Euclidean wormhole in terms of boundary degrees of freedom.

We begin with 3d gravity in the first order formulation and continue to Euclidean signature. The choice of continuation made in~\cite{Cotler:2020ugk}, made mostly in order to ensure that fields that appear linearly in the Euclidean action act as Lagrange multipliers, was to simply rotate time as $t=-i y$ and rotate the contours of time components of the dreibein and spin connection, i.e.
\beq
	e^A_t = i e^A_y\,, \qquad \omega^A{}_{Bt} = i \omega^A{}_{By}\,,
\eeq 
and to integrate over real contours for all fields. In particular the approach of~\cite{Cotler:2020ugk} did not continue the flat Minkowski metric $\eta_{AB}$ with a flat Euclidean metric $\delta_{AB}$.

After this continuation we explicitly separate the Euclidean time components from the spatial ones through $A^A = A^A_y dy + A^A_i dx^i$, so that the 3d gravity action reads
\begin{equation}\label{EuclideanGravityCSAction}
    S = -\frac{ik}{4 \pi} \int dyd^2\,\epsilon^{ij} tr \left(-A_{i} \partial_{y} A_{j} + A_{y} F_{ij} \right) - (A \rightarrow \bar{A}) + S_{\rm bdy},\,.
\end{equation}
In~\cite{} the authors worked in coordinates $x^i = (x,\rho)$ with boundary coordinates $x\sim x+2\pi$ and $y\sim y+2\pi$, while $\rho\in \mathbb{R}$ is a radial coordinate, interpolating between a torus conformal boundary with complex structure $\tau_2$ as $\rho\to-\infty$ and a torus conformal boundary with complex structure $\tau_1$ as $\rho\to\infty$. These boundary conditions 
amount to the statements\footnote{Here we have made the choice $J_0=-\frac{i\sigma_2}{2}$, $J_1=\frac{\sigma_1}{2}$, and $J_2 = \frac{\sigma_3}{2}$, and represented the $\bar{J}_A$ in the same way.}
\begin{equation}
\rho \rightarrow \infty: \, \,    A \approx \frac{1}{2}
    \begin{pmatrix}
    d \rho & 0 \\
    e^{\rho} (dx + \bar{\tau}_{1}dy) & - d \rho
    \end{pmatrix}
    , \quad  \bar{A} \approx \frac{1}{2}
    \begin{pmatrix}
    -d \rho & -e^{\rho} (dx + \tau_{1}dy) \\
    0 &  d \rho
    \end{pmatrix}
    \,,
\end{equation}
and
\begin{equation}
\rho \rightarrow -\infty: \, \,    A \approx \frac{1}{2}
    \begin{pmatrix}
    d \rho & e^{-\rho} (dx + \bar{\tau}_{2}dy) \\
    0 & - d \rho
    \end{pmatrix}
    , \quad  \bar{A} \approx \frac{1}{2}
    \begin{pmatrix}
    -d \rho & 0 \\
    -e^{-\rho} (dx + \tau_{2}dy) &  d \rho
    \end{pmatrix}
    \,,
\end{equation}
so that the line element is asymptotically
\begin{equation}
    ds^{2} \approx \frac{e^{2 \rho}}{4} |dx + \tau_{1,2} dy|^{2} + d \rho^{2}\,.
\end{equation}
The boundary term $S_{\rm bdy}$ is fixed by the requirement that there is a consistent variational principle with those boundary conditions, and reads
\begin{equation}\label{BoundaryPartofEGA}
    S_{bdy} = \frac{ik}{4 \pi} \left(\int_{\rho \rightarrow \infty} d^{2} x\, \text{tr} \left(\bar{\tau}_{1} A^{2}_{x} - \tau_{1} \bar{A}^{2}_{x} \right) + \int_{\rho \rightarrow -\infty} d^{2} x \,\text{tr} \left(\bar{\tau}_{2} A^{2}_{x} - \tau_{2} \bar{A}^{2}_{x} \right) \right)\,.
\end{equation}

The temporal component $A_y$ appears linearly in the action and so acts as a Lagrange multiplier enforcing that the spatial field strength vanishes, $\epsilon^{ij}F_{ij} = 0$. Integrating it out, we are left with a residual integral over those $A_i$ satisfying the constraint, which can be parameterized as
\begin{equation}\label{SpatialCSTerms}
    A_i = g^{-1} \partial_i g\,, \quad \bar{A}_i = \bar{g}^{-1} \partial_i\bar{g}\,,
\end{equation}
for some matrix-valued functions $g$ and $\bar{g}$. Doing so introduces a redundancy under time-dependent $SL(2;\mathbb{R})\times SL(2;\mathbb{R})$ transformations, which appears as a gauge symmetry and acts as $g(x^i,y) \to h(y) g(x^i,y)$ and $\bar{g} (x^i,y) \to \bar{h}(y) \bar{g}(x^i,y)$. 

Because the constant time slice is an annulus, these functions need not be single-valued around the $x$-circle. We can use the $SL(2;\mathbb{R})\times SL(2;\mathbb{R})$ freedom to fix $g$ and $\bar{g}$ to take the form 
\begin{equation}
     g = e^{b(y)x J_{1}} \widetilde{g}\,, \quad \bar{g} = e^{\bar{b}(y)x \bar{J}_{1}} \widetilde{\bar{g}}\,,
\end{equation}
where $\widetilde{g}$ and $\widetilde{\bar{g}}$ are single-valued around the $x$-circle. This partial gauge-fixing leaves behind a redundancy under those $SL(2;\mathbb{R})\times SL(2;\mathbb{R})$ transformations that commute with the prefactors $e^{b(y) x J_1}$ and $e^{\bar{b}(y)x\bar{J}_1}$, namely a $U(1)\times U(1)$ redundancy under
\begin{equation}
    h = e^{a(y)J_{1}}\,, \quad \bar{h} = e^{\bar{a}(y) \bar{J}_{1}}\,.
\end{equation}
If we were quantizing Chern-Simons theory rather than gravity, then $b(y)$ and $\bar{b}(y)$ would parameterize time-dependent holonomies for $A$ and $\bar{A}$ around the $x$-circle. 

The gravity action evaluated on these configurations is a pure boundary term. At this stage there are six real boundary degrees of freedom on each boundary torus, three from $\widetilde{g}$ and three from $\widetilde{\bar{g}}$, in addition to the two ``quantum mechanical'' degrees of freedom $b$ and $\bar{b}$. On each boundary the asymptotically AdS$_3$ boundary conditions fix four of the six degrees of freedom in $\widetilde{g}$ and $\widetilde{\bar{g}}$. In more detail, it is convenient to decompose $\tilde{g}$ and $\tilde{\bar{g}}$ as
\beq
	\tilde{g} = e^{\phi J_1} e^{\Lambda J_2} e^{\psi(J_1-J_0)}\,, \qquad \tilde{\bar{g}} = e^{\bar{\phi}\bar{J}_1} e^{-\bar{\Lambda}\bar{J}_2} e^{\bar{\psi}(\bar{J}_1+\bar{J}_0)}\,,
\eeq
so that the fields $\phi$ and $b$ (and $\bar{\phi}$ and $\bar{b}$) appear in the combinations
\beq
	\Phi(x,y,\rho) = b(y)x + \phi(x,y,\rho)\,,  \qquad \bar{\Phi}(x,y,\rho) = \bar{b}(y)x + \bar{\phi}(x,y,\rho)\,.
\eeq
The $U(1)\times U(1)$ gauge symmetry means that we identify
\beq
\label{E:U1gauge}
	\phi(x,y,\rho) \sim \phi(x,y,\rho)+a(y)\,, \qquad \bar{\phi}(x,y,\rho) \sim \bar{\phi}(x,y,\rho) + \bar{a}(y)\,.
\eeq
The asymptotically AdS boundary conditions imply that, at large $\rho$, the fields $\Lambda$ and $\psi$ are fixed in terms of $\Phi$, and $\bar{\Lambda}$ and $\bar{\psi}$ in terms of $\bar{\Phi}$ as
\beq
	\Lambda \approx \ln\left( \frac{e^{\rho}}{\Phi'}\right)\,, \qquad \psi\approx - \frac{e^{-\rho}\Phi''}{\Phi'}\,, \qquad \bar{\Lambda} \approx \ln\left( \frac{e^{\rho}}{\bar{\Phi}'}\right)\,, \qquad \bar{\psi} \approx - \frac{e^{-\rho}\bar{\Phi}''}{\bar{\Phi}'}\,,
\eeq
where $' = \partial_x$. There are similar expressions near the other boundary as $\rho\to-\infty$. We denote the boundary values of $\Phi$ and $\bar{\Phi}$ as $\Phi_1 = \lim_{\rho\to\infty} \Phi$, $\bar{\Phi}_1 = \lim_{\rho\to\infty} \bar{\Phi}$, $\Phi_2 = \lim_{\rho\to-\infty} \Phi$, and $\bar{\Phi}_2 = \lim_{\rho\to-\infty} \Phi$. 

Evaluating the gravity action~\eqref{EuclideanGravityCSAction} including the boundary term~\eqref{BoundaryPartofEGA} on these configurations leads to the Euclidean boundary action\footnote{This corrects some misprints present in \cite{Cotler:2020ugk}.}
\begin{align}
\begin{split}
\label{EuclideanWormhole}
	S_{E}& = \frac{C}{24 \pi} \int d^{2}x \bigg(\frac{\Phi''_{1}\partial_{1} \Phi'_{1}}{\Phi'^{2}_{1}} + \frac{\bar{\Phi}''_{1} \bar{\partial}_{1} \bar{\Phi}'_{1}}{\bar{\Phi}'^{2}_{1}} + \frac{i}{2} \left(\bar{\tau}_{1} \Phi'^{2}_{1} - \phi'_{1}\partial_{y} \phi_{1} - \tau_{1} \bar{\Phi}'^{2}_{1} + \bar{\phi}'_{1} \partial_{y} \bar{\phi}_{1} \right) 
	\\
	   & \qquad \qquad \qquad \qquad +\frac{\Phi''_{2}\partial_{2} \Phi'_{2}}{\Phi'^{2}_{2}} + \frac{\bar{\Phi}''_{2} \bar{\partial}_{2} \bar{\Phi}'_{2}}{\bar{\Phi}'^{2}_{2}} + \frac{i}{2} \left(\bar{\tau}_{2} \Phi'^{2}_{2} + \phi'_{2}\partial_{y} \phi_{2} - \tau_{2} \bar{\Phi}'^{2}_{2} - \bar{\phi}'_{2} \partial_{y} \bar{\phi}_{2} \right) \bigg)
	\\
	& \qquad \qquad \qquad \qquad \qquad- \frac{iC}{24} \int^{2 \pi}_{0} dy \, \left(b^{2} \partial_{y} Y - \bar{b}^{2} \partial_{y} \bar{Y} \right), 
\end{split}
\end{align}
where $C=\frac{3}{2G}$, the derivatives are defined as
\begin{equation}
    \partial_{1} = \frac{i}{2} (\bar{\tau}_{1} \partial_{x} - \partial_{y}), \quad \quad \partial_{2} = \frac{i}{2} (\bar{\tau}_{2} \partial_{x} + \partial_{y})\,,
\end{equation}
and the fields $Y$ and $\bar{Y}$ are defined as
\begin{equation}
    Y(y) = \frac{1}{2 \pi b(y)} \int^{2 \pi}_{0} dx \, \left(\phi_{1}(x,y) - \phi_{2}(x,y) \right) \,, \quad    \bar{Y}(y) = \frac{1}{2 \pi \bar{b}(y)} \int^{2 \pi}_{0} dx \, \left(\bar{\phi}_{1}(x,y) - \bar{\phi}_{2}(x,y) \right).
\end{equation}
Note that this action is indeed invariant under the $U(1)\times U(1)$ gauge transformations in~\eqref{E:U1gauge} . It has a single time derivative and therefore is in Hamiltonian form.

Let us briefly comment on the quantum mechanical treatment of this model in~\cite{Cotler:2020ugk}. One can treat the ``twist fields'' $Y$ and $\bar{Y}$ as independent quantum mechanical degrees of freedom at the cost of introducing a further redundancy under which $\phi_i \sim \phi_i + a_i(y)$ and $\bar{\phi}_i \sim \bar{\phi}_i + \bar{a}_i(y)$ with $i=1,2$. The twist fields then only appear in the last line of the action. Integrating them out enforces that the fields $b$ and $\bar{b}$ are constant moduli which one integrates over. Requiring that the spatial geometry is smooth enforces that $b^2$ and $\bar{b}^2$ are both non-negative. At fixed moduli $b$ and $\bar{b}$ one can find a saddle point for the remaining fields $\phi_i$ and $\bar{\phi}_i$ (where they simply vanish), and in fact the path integral over those fields is one-loop exact. However the action varies in the directions of $b$ and $\bar{b}$, so that this model does not have a true saddle point, consistent with the fact that three-dimensional gravity lacks a wormhole saddle with this topology. It is helpful to think of $b$ and $\bar{b}$ as moduli that have to be stabilized to find a genuine saddle.

\subsection{Correlation functions and Wilson lines}

We would now like to understand the effect of a Gao-Jafferis-Wall deformation as a deformation of the action~\eqref{EuclideanWormhole}. As we mentioned in the Introduction, we can replace the effect of the large number of matter fields and double trace deformation by an insertion of $\exp\left( \sum_{i=1}^{N_f}\int dt dx \,g\langle O_L^i(t,x)O_R^i(t,x)\rangle\right)$ into the sum over metrics with $\langle O_L^i O_R^i\rangle$ the left-right boundary-to-boundary two-point function of $O$ across the wormhole. So, to proceed, we need the left-right two-point function of a  operator $O$ across the Euclidean wormholes in the previous Subsection, in terms of the gravitational degrees of freedom; see Eq.~\eqref{E:wilsonLine} for the result; and to then analytically continue to Lorentzian signature.

We compute $\langle O_L O_R\rangle$ where the two insertions are at the same $y$ for a fixed wormhole by considering a Wilson line in the gauge theory description stretching from $(x_2,y,\rho\to -\infty)$ to $(x_1,y,\rho\to +\infty)$. For a scalar field this Wilson line simply encodes the renormalized geodesic length between these two points. Because the constant$-y$ slice is an annulus, Wilson lines from one boundary to the other can wind around the $x$-circle, and we sum over all such windings.

As we mentioned at the end of the previous Subsection, at fixed $b$ and $\bar{b}$, there is a Euclidean wormhole saddle with $\phi_i = \bar{\phi}_i=0$. For this wormhole we can reconstruct the spatial part of the gauge configuration $A_i$ and $\bar{A}_j$, and so the spatial metric and spin connection. The result is
\beq
\label{E:representative}
	A_idx^i = \frac{1}{2}\begin{pmatrix} d\rho  & b e^{-\rho} dx \\ b e^{\rho} dx & - d\rho\end{pmatrix} \,, \qquad \bar{A}_i dx^i = \frac{1}{2}\begin{pmatrix} - d\rho & - \bar{b} e^{\rho}dx \\ - \bar{b} e^{-\rho} dx & d\rho\end{pmatrix}\,,
\eeq
or equivalently
\beq
	g = e^{b x J_0} e^{\rho J_2}\,, \qquad \bar{g} = e^{\bar{b}x \bar{J}_0} e^{-\rho \bar{J}_2}\,.
\eeq
The holographically renormalized Wilson line in the principal unitary irreducible representation with scaling weights $(h,\bar{h})$~\cite{Ammon:2013hba} that does not wind around the $x$-circle is
\beq
\label{E:basicWilson}
	W_{h,\bar{h}}(x_1,x_2) =b^{2h}\bar{b}^{2\bar{h}} \left( 2 \cosh\left( \frac{b x_{12}}{2}\right)\right)^{-2h}\left( 2 \cosh\left( \frac{\bar{b}x_{12}}{2}\right)\right)^{-2\bar{h}}\,. 
\eeq
Allowing for a general configuration of boundary graviton degrees of freedom effectively reparameterizes this quantity, and summing over all possible windings around the $x$-circle we have
\beq
\label{E:wilsonLine}
	\mathcal{W}_{h,\bar{h}}(x_1,x_2) =\sum_{n=-\infty}^{\infty} \left( \frac{\Phi_1'(x_1)\Phi_2'(x_2)}{\left( 2 \cosh\left( \frac{b(x_{12}+2\pi n)+\phi_{12}}{2}\right)\right)^2}\right)^{h} \left( \frac{\bar{\Phi}_1'(x_1) \bar{\Phi}_2'(x_2)}{\left( 2 \cosh\left( \frac{\bar{b}(x_{12}+2\pi n)+\bar{\phi}_{12}}{2}\right)\right)^2}\right)^{\bar{h}}\,,
\eeq
where $x_{12}=x_1-x_2$, $\phi_{12} = \phi_1(x_1)-\phi_2(x_2)$, $\bar{\phi}_{12}=\bar{\phi}_1(x_1)-\bar{\phi}_2(x_2)$, and all fields are implicitly at the same time $y$. Note that the sum over windings renders the Wilson line single-valued around the $x$-circle. As an operator it is also single-valued in $y$ as it is a function of fields periodic in $y$.\footnote{In the boundary graviton theory there are, at fixed $b$ and $\bar{b}$, saddles where the $\phi_i$ and $\bar{\phi}_j$ wind around the $y$-circle. We expect that those saddles contribute to the expectation value of $\mathcal{W}_{h,\bar{h}}$ in the boundary theory, in such a way that the result is single-valued both in $x$ and $y$.} We propose that, to leading order as $G\to 0$, this quantity is proportional to $\langle O_L O_R\rangle$ where the boundary operator $O$ carries scaling weights $(h,\bar{h})$. In the next Section we will see that, upon a suitable continuation to real time, this proposal can also be derived from other considerations. For now we note that it takes the form of a bilocal operator in the boundary Alekseev-Shatashvili theory~\cite{Cotler:2018zff}, a reparameterized two-point function, similar to the bilocals appearing in the Schwarzian model governing the edge modes of JT gravity.

Before going on we observe that for $h=\bar{h}$, i.e. when $O$ is a scalar of dimension $\Delta = h+\bar{h}$, the basic Wilson line~\eqref{E:basicWilson} can also obtained from a geodesic in the spatial geometry that follows from~\eqref{E:representative},
\beq
	ds_{\rm spatial}^2 = d\rho^2 + \left( b \bar{b}\sinh^2(\rho) + \frac{(b+\bar{b})^2}{4}\right)dx^2\,.
\eeq
For simplicity take $b=\bar{b}$ so that $ds^2_{\rm spatial} = d\rho^2 + b^2 \cosh^2(\rho) dx^2$. Parameterizing the path as $x=x(\rho)$, the point particle action $\int d\rho \sqrt{1 + b^2 \cosh^2\rho\,x'(\rho)^2}$ has a constant of the motion $p=\frac{\partial L}{\partial x'(\rho)}$ that can be traded for $x_{12}$ with the result $p = b \tanh\left(\frac{b x_{12}}{2}\right)$. The renormalized geodesic length is
\beq
	L = 2 \ln \left(\frac{\left( 2 \cosh\left( \frac{b x_{12}}{2}\right)\right)^2}{b^2}\right)\,.
\eeq
This is the length of the geodesic that does not wind around the $x-$circle. The ones that do have the same length with the replacement $x_{12} \to x_{12} + 2\pi n$. The geometric optics approximation to $\langle O_L O_R\rangle$ is then
\beq
	\langle O_L(x_2) O_R(x_1) \rangle \sim \sum_n e^{-\Delta L} =\sum_{n=-\infty}^{\infty} \left( \frac{b^2}{\left( 2 \cosh\left( \frac{b (x_{12}+2\pi n)}{2}\right)\right)^2}\right)^{h + \bar{h}}\,.
\eeq
The geometric optics approximation is just that, an approximation, valid at large $h+\bar{h}$. However soon we will see that the general expression~\eqref{E:wilsonLine} receives no corrections to leading order in the gravitational interaction.

\section{Traversable wormholes in 3d}
\label{S:wormholes}
\subsection{Continuing to real time}
\label{S:wormholeAction}

In the previous Section we reviewed the action obtained in~\cite{Cotler:2020ugk} describing Euclidean wormholes in pure three-dimensional gravity on spaces of the topology $\mathbb{T}^2\times \mathbb{R}$, and found the semiclassical approximation to the left-right two-point function $\langle O_L O_R\rangle$ across the wormhole. See Eqs.~\eqref{EuclideanWormhole} and~\eqref{E:wilsonLine}. Crucially the Euclidean wormhole action did not possess a saddle point. The parameters $b$ and $\bar{b}$ appearing there are moduli which have to be stabilized to find a genuine saddle. Now we analytically continue to real time, add the Gao-Jafferis-Wall deformation, and search for wormhole saddles.

In Euclidean signature the space at constant $\rho$ is a torus parameterized by circles $x$ and $y$. In the derivation of the boundary action~\eqref{EuclideanWormhole} it was convenient to think of $y$ as Euclidean time and $x$ as space, but in what follows we find it convenient to flip roles, thinking of what was the $x$-circle as Euclidean time, which we will Wick-rotate back to Lorentzian signature. To do so we have to perform some relabeling gymnastics, since we would like to still use the label $x$ as a coordinate on the spatial circle. Our first step, in Euclidean signature, is to exchange $x$ and $y$:
\beq
	(x,y) \to (y,x)\,.
\eeq
In so doing the ``quantum mechanical'' degrees of freedom $b, \bar{b}, Y,$ and $\bar{Y}$ become functions of $x$ alone, so that e.g. $\Phi_i = b(x) y + \phi_i(x,y)$. Since we are interested in a Lorentzian configuration where the new spatial circle $x$ is periodic with $x\sim x+2\pi$ and time $t=-iy$ is non-compact, we first decompactify the Euclidean torus. This can be done by setting $\tau_1=\tau_2=i$ and then letting $y\in \mathbb{R}$. We also continue $\phi_i \to i \phi_i$ and $\bar{\phi}_i \to i \bar{\phi}_i$ so that the real-time fields
\beq
	\Phi_i(t,x) = b(x)t + \phi_i(t,x)\,, \qquad \bar{\Phi}_i(t,x) = \bar{b}(x)t + \bar{\phi}_i(t,x)\,,
\eeq
are real. The real-time boundary action so obtained is then
\begin{align}
\begin{split}
\label{E:lorentzianS}
	S_L & = \frac{C}{24\pi}\int dt dx \left( \frac{\ddot{\Phi}_1 \partial_+ \dot{\Phi}_1}{\dot{\Phi}_1^2} + \frac{\ddot{\bar{\Phi}}_1 \partial_- \dot{\bar{\Phi}}_1}{\dot{\bar{\Phi}}_1^2} - \frac{1}{2}\left( \dot{\Phi}_1^2 + \dot{\phi}_1 \phi_1' + \dot{\bar{\Phi}}_1^2 - \dot{\bar{\phi}}_1 \bar{\phi}_1' \right) \right.
	 \\
	 & \qquad \qquad \qquad \qquad+ \left.\frac{\ddot{\Phi}_2\partial_- \dot{\Phi}_2}{\dot{\Phi}_2^2} + \frac{\ddot{\bar{\Phi}}_2 \partial_+ \dot{\bar{\Phi}}_2}{\dot{\bar{\Phi}}_2^2} - \frac{1}{2}\left( \dot{\Phi}_2^2 - \dot{\phi}_2 \phi_2' + \dot{\bar{\Phi}}_2^2 + \dot{\bar{\phi}}_2 \bar{\phi}_2'\right)\right)
	 \\
	 & \qquad \qquad \qquad\qquad \qquad- \frac{C}{24}\int_0^{2\pi} dx\,\left( b^2 Y' - \bar{b}^2 \bar{Y}'\right)\,,
\end{split}
\end{align}
where $\dot{\phantom{.}}=\partial_t$, $'=\partial_x$, $\partial_{\pm} = \frac{1}{2}(\partial_t\pm \partial_x)$, and
\beq
	Y(x) = \frac{1}{2\pi b(x)} \int dt\left( \phi_1(t,x)-\phi_2(t,x)\right)\,, \qquad \bar{Y}(x) = \frac{1}{2\pi\bar{b}(x)}\int dt \left( \bar{\phi}_1(t,x) - \bar{\phi}_2(t,x)\right)\,.
\eeq
The left-right two-point function~\eqref{E:wilsonLine} however becomes 
\beq
\label{E:doubleTraceL}
	\langle O_L(t_2,x) O_R(t_1,x)\rangle =  \left( \frac{\dot{\Phi}_1(t_1)\dot{\Phi}_2(t_2)}{\left( 2\phantom{\frac{\bar{\Phi}}{2}}\hspace{-.12in} \cos\left( \frac{\Phi_{12}}{2}\right)\right)^2}\right)^h\left( \frac{\dot{\bar{\Phi}}_1(t_1)\dot{\bar{\Phi}}_2(t_2)}{\left( 2 \cos\left( \frac{\bar{\Phi}_{12}}{2}\right)\right)^2} \right)^{\bar{h}}\,,
\eeq
where $\Phi_{12} = \Phi_1(t_1)-\Phi_2(t_2)$, $\bar{\Phi}_{12}=\bar{\Phi}_1(t_1)-\bar{\Phi}_2(t_2)$, and all fields are evaluated at the same spatial position $x$. Note that, unlike~\eqref{E:wilsonLine}, this expression has no additional images; upon trading $x$ and $y$ and letting the $y$ circle have a radius $L$, the images in~\eqref{E:wilsonLine} are suppressed by powers of $\text{sech}(\pi b L)$ and $\text{sech}(\pi \bar{b}L)$ and so vanish as we decompactify Euclidean time. As an operator in the boundary graviton theory this bilocal is single-valued around the $x$-circle, as it depends on fields which are all periodic in $x$. We expect that its expectation value is rendered single-valued thanks to saddles where the $\phi_i$ and $\bar{\phi}_j$ wind around the $x$-circle.

In the Euclidean problem we had a $U(1)\times U(1)$ gauge symmetry identifying $\phi_i(x,y) \sim \phi_i(x,y) + a(x)$ and $\bar{\phi}_i(x,y)\sim \bar{\phi}_i(x,y) + \bar{a}(x)$ (after exchanging $x\leftrightarrow y$). The boundary graviton action~\eqref{E:lorentzianS} and reparameterized two-point function~\eqref{E:doubleTraceL} are in fact invariant under a larger $SO(2,2)$ gauge symmetry that arises after decompactifying $y$. (A finite circle size can be thought of as breaking $SO(2,2)$ to $U(1)\times U(1)$.) It acts as
\begin{align}
\begin{split}
	\tan\left( \frac{\Phi_1}{2}\right) &\sim \frac{\alpha\tan\left( \frac{\Phi_1}{2}\right) + \beta}{\gamma \tan\left( \frac{\Phi_1}{2}\right)+\delta} \,, \qquad \,\cot\left( \frac{\Phi_2}{2}\right) \sim \frac{\alpha \cot\left( \frac{\Phi_2}{2}\right) - \beta }{-\gamma\cot\left( \frac{\Phi_2}{2}\right)+\delta}\,,
	\\
	\tan\left( \frac{\bar{\Phi}_1}{2}\right) & \sim \frac{\bar{\alpha} \tan\left(\frac{ \bar{\Phi}_1}{2}\right)+\bar{\beta}}{\bar{\gamma} \tan\left( \frac{\bar{\Phi}_1}{2}\right)+\bar{\delta}}\,, \qquad \cot\left( \frac{\bar{\Phi}_2}{2}\right) \sim \frac{\bar{\alpha}\cot\left( \frac{\bar{\Phi}_2}{2}\right)-\bar{\beta}}{-\bar{\gamma}\cot\left( \frac{\bar{\Phi}_2}{2}\right)+\bar{\delta}}\,,
\end{split}
\end{align}
where $\alpha \delta - \beta \gamma=\bar{\alpha}\bar{\delta}-\bar{\beta}\bar{\gamma}=1$ and all of these parameters are periodic functions of $x$.

In fact, the reparameterized two-point function is the unique gauge-invariant operator with the right scaling behavior that can couple the two boundaries together. This gives a complementary argument for how the double trace deforms the boundary graviton theory.

We have taken a bit of a roundabout path to arrive at the Lorentzian action~\eqref{E:lorentzianS} and correlator~\eqref{E:doubleTraceL}, taking a detour through Euclidean signature and performing a double Wick rotation along the way. A more direct route is, in Lorentzian signature, to first integrate out $A_x$. This approach was taken in Section 4 of~\cite{Cotler:2018zff} to describe a boundary action for the two-sided BTZ black hole, but it is actually quite tricky to get the precise field ranges and action correct. Indeed the action obtained there was almost, but not quite correct. It is for this reason that we took a long detour through imaginary time.

We reap two benefits as well. First, the dressed two-point function~\eqref{E:doubleTraceL} has no sum over images. Second, the action obtained here is conceptually quite similar to that obtained by Maldacena and Qi in their study of traversable wormholes in JT gravity. Indeed, in a sense the JT traversable wormhole is the analytic continuation of a two-sided black hole, and roughly speaking the boundary action obtained here is the analytic continuation of that of a two-sided BTZ black hole.

Adding an eternal version of the Gao-Jafferis-Wall deformation we arrive at the total action
\beq
\label{E:totalS}
	S_{\rm total} = S_L + S_{DT}\,, \qquad S_{DT} = \frac{C}{24\pi} \int dt dx \,\eta \left( \frac{\dot{\Phi}_1(t_1)\dot{\Phi}_2(t_2)}{\phantom{\left(\frac{\bar{\Phi}}{2}\right)}\hspace{-.27in}\cos^2\left( \frac{\Phi_{12}}{2}\right)}\right)^{h}\left( \frac{\dot{\bar{\Phi}}_1(t_1)\dot{\bar{\Phi}}_2(t_2)}{\cos^2\left( \frac{\bar{\Phi}_{12}}{2}\right)} \right)^{\bar{h}}\,,
\eeq
where $\eta =O(1)>0$ controls the strength of the double trace deformation. In the remainder of this Section we demonstrate that this action has saddles corresponding to stable wormholes for $h+\bar{h}<1$.

\subsection{Wormhole saddles and fluctuation spectrum}

We wish to find translationally invariant solutions to the field equations of the deformed action~\eqref{E:totalS}, characterized by the ansatz
\begin{equation}\label{fieldansatz}
    \phi_{1} = \phi_0\,,\bar{\phi}_{1} =\bar{\phi}_0\,, \qquad \phi_2= \bar{\phi}_{2} = 0\,, \qquad b(t) = b\,, \qquad \bar{b}(t) = \bar{b}\,.
\end{equation}
The Lagrangian evaluated on this ansatz is simply
\beq
	\mathcal{L} = \frac{C}{24\pi}\left( -b^2 - \bar{b}^2 + \eta \frac{b^{2h}\bar{b}^{2\bar{h}}}{\cos^{2h}\left( \frac{\phi_0}{2}\right) \cos^{2\bar{h}}\left( \frac{\bar{\phi}_0}{2}\right)} \right)\,.
\eeq
Extremizing with respect to $\phi_0$ and $\bar{\phi}_0$ we find 
\beq
	\phi_0 = \bar{\phi}_0=0\,,
\eeq
while extremizing with respect to $b$ and $\bar{b}$ fixes
\begin{align}
\begin{split}
\label{E:bSaddle}
    b &= \left(2\, (\Delta - s)^{\left(\frac{s - \Delta}{2}\right)} \, (\Delta + s)^{\left(\frac{\Delta - s }{2}- 1\right)} \, \eta^{-1} \right)^{\frac{1}{2(\Delta -1)} }\,,
    \\
   \bar{b} &= \left(2\, (\Delta - s)^{\left(\frac{s+ \Delta }{2}- 1\right)}  \, (\Delta + s)^{-\left(\frac{s+\Delta}{2}\right)} \, \eta^{-1} \right)^{\frac{1}{2(\Delta -1)} }\,,
\end{split}
\end{align}
where we have traded $h$ and $\bar{h}$ for $\Delta = h+\bar{h}$ and $s=h-\bar{h}$. One can verify that this configuration is a solution to the full equations of motion of the model.

As advertised, the double trace deformation stabilizes the moduli leading to a saddle point which, in three dimensions, is a traversable wormhole. Due to the exponent $\frac{1}{2(\Delta-1)}$ the saddle behaves as $b,\bar{b}\to 0$ at fixed deformation $\eta$ as $\Delta \to 1$ from below, i.e. the saddle becomes singular in the limit that the double trace deformation goes from being relevant to marginal. Indeed, this is a precursor to the fact that this saddle is stable only when $\Delta<1$.

Next we study the spectrum of fluctuations around this saddle. Doing so is a little tricky, since some of the fields depend only on the spatial circle $x$, while the others depend on time as well.

Before going on we note that the saddle point values for $b$ and $\bar{b}$ satisfy two relations which are useful in simplifying the fluctuation problem, namely
\beq
\label{E:useful}
	\eta = \frac{b^{1-2h} \bar{b}^{1-2\bar{h}}}{\sqrt{h \bar{h}}}\,, \qquad b^2 \bar{h} = \bar{b}^2 h\,.
\eeq

Now consider a general fluctuation characterized by
\begin{align}
\begin{split}
	\Phi_1 & = (b+\delta b(x))t + \phi_1(t,x)\,, \qquad \qquad \Phi_2 = (b+\delta b(x))t + \phi_2(t,x) \,,
	\\
	\bar{\Phi}_1 & = (\bar{b}+\delta\bar{b}(x))t + \bar{\phi}_1(t,x) \,, \qquad \qquad \bar{\Phi}_2 = (\bar{b}+\delta\bar{b}(x))t+\bar{\phi}_2(t,x)\,,
\end{split}
\end{align}
Fourier transforming the fluctuations we have
\begin{align}
\begin{split}
	\phi_i(t,x) & = \frac{1}{(2\pi)^2}\sum_{k\in\mathbb{Z}}\int d\omega \,e^{-i \omega t + i k x} \phi_i(k,\omega)\,,
	\\
	\bar{\phi}_i(t,x) & = \frac{1}{(2\pi)^2} \sum_{k\in\mathbb{Z}}\int d\omega \,e^{-i\omega t + i k x}\bar{\phi}_i(k,\omega)\,,
	\\
	\delta b(x) & = \frac{1}{2\pi}\sum_{k\in\mathbb{Z}} e^{ikx} \delta b(k)\,,
	\\
	\delta\bar{b}(x) & = \frac{1}{2\pi}\sum_{k\in\mathbb{Z}}e^{ikx}\delta\bar{b}(k)\,.
\end{split}
\end{align}

Now we must take into account the $SO(2,2)$ gauge symmetry. Infinitesimal gauge transformations act on the perturbations as
\begin{align}
\begin{split}
	\delta \Phi_1 &= e^{i b t} c_+(x) + c_0(x)+ e^{-i b t} c_-(x) \,,
	\\
	\delta \Phi_2 & = -e^{i b t} c_+(x) + c_0(x)- e^{-i b t}c_-(x) \,,
	\\
	\delta\bar{\Phi}_1 & = e^{i \bar{b}t} \bar{c}_+(x) + \bar{c}_0(x)+ e^{-i \bar{b}t} \bar{c}_-(x) \,,
	\\
	\delta\bar{\Phi}_2 & = -e^{i \bar{b}t} \bar{c}_+(x)+ \bar{c}_0(x)- e^{-i \bar{b}t} \bar{c}_-(x) \,.
\end{split}
\end{align}
None of these terms are linear in $t$ and so $\delta b$ and $\delta\bar{b}$ are inert under them. The terms involving $c_0$ and $\bar{c}_0$ generate the original $U(1)\times U(1)$ symmetry, while those involving $c_{\pm}$ and $\bar{c}_{\pm}$ fill out the rest of $SO(2,2)$. The former act only the $\omega=0$ modes of the $\phi_i$ and $\bar{\phi}_i$, while the latter act on the $\omega = \pm b$ and $\omega = \pm \bar{b}$ modes. The $U(1)\times U(1)$ subgroup can be used to set $\phi_2(k,\omega=0)$ and $\bar{\phi}_2(k,\omega=0)$ to vanish, and we redefine
\beq
	\phi_1(k,\omega) \to \phi_1(k,\omega) + 2\pi \delta(\omega) \mathcal{Y}(k)\,, \qquad \bar{\phi}_1(k,\omega)\to \bar{\phi}_1(k,\omega) + 2\pi \delta(\omega)\bar{\mathcal{Y}}(k)\,,
\eeq
where the new fields $\phi_1(k,\omega)$ and $\bar{\phi}_1(k,\omega)$ vanish at $\omega=0$. The $\omega=0$ fields are $\delta b, \delta\bar{b}, \mathcal{Y}$, and $\bar{\mathcal{Y}}$. We treat the $\omega = \pm b$ and $\omega = \pm \bar{b}$ modes similarly.

\subsubsection{$\omega\neq 0$}
We begin with the modes carrying generic nonzero frequency, $\phi_i(k,\omega)$ and $\bar{\phi}_i(k,\omega)$. The quadratic action for these modes is 
\beq
	S_{\rm quad} = \frac{1}{(2\pi)^2}\sum_{k\in\mathbb{Z}} \int d\omega \,\frac{1}{2}\mathbf{\Phi}^{\dagger}\cdot G^{-1}\mathbf{\Phi}\,,
\eeq

where 
\beq
\label{E:bdyGravitonPropagator}
	\mathbf{\Phi} = \begin{pmatrix} \phi_1(k,\omega) \\ \phi_2(k,\omega) \\ \bar{\phi}_1(k,\omega) \\ \bar{\phi}_2(k,\omega) \end{pmatrix}\,, \qquad G^{-1} = \frac{C}{24\pi} \begin{pmatrix} \chi(k,\omega) & -\frac{b^2}{2}+h\omega^2 & \sqrt{h \bar{h}}\omega^2 & \sqrt{h\bar{h}}\omega^2 \\ -\frac{b^2}{2} + h \omega^2 & \chi(k,-\omega) & \sqrt{h \bar{h}}\omega^2 & \sqrt{h\bar{h}}\omega^2 \\  \sqrt{h \bar{h}}\omega^2 &  \sqrt{h \bar{h}}\omega^2 & \bar{\chi}(k,-\omega) & -\frac{\bar{b}^2}{2} + \bar{h}\omega^2 \\  \sqrt{h \bar{h}}\omega^2 &  \sqrt{h \bar{h}}\omega^2 & -\frac{\bar{b}^2}{2} + \bar{h}\omega^2 & \bar{\chi}(k,\omega) \end{pmatrix}\,,
\eeq
and we have defined
\begin{align}
\begin{split}
	\chi(k,\omega) &= \frac{b^2}{2} +\omega^2(h-1) + \frac{\omega(b^2-\omega^2)(k-\omega)}{b^2}\,,
	\\
	\bar{\chi}(k,\omega) & =\frac{\bar{b}^2}{2} +\omega^2(\bar{h}-1) + \frac{\omega(\bar{b}^2-\omega^2)(k-\omega)}{\bar{b}^2}\,.
\end{split}
\end{align}
The linearized equations of motion are simply
\beq
	G^{-1}\mathbf{\Phi}=\mathbf{0}\,.
\eeq
As usual solutions exist only for those values of $\omega$ and $k$ such that $G^{-1}$ has a zero-eigenvalue. To find them we solve $\text{det}\,G^{-1}=0$. That determinant is
\beq
\label{E:det}
	\text{det}\,G^{-1} =\left( \frac{C}{24\pi}\right)^4 \frac{\omega^4(\omega^2-b^2)^2(\omega^2-\bar{b}^2)^2}{b^4\bar{b}^4} \left(  (\omega^2-k^2)-m_+^2)((\omega^2-k^2)-m_-^2\right)\,,
\eeq
with
\beq
\label{E:massSquared}
	m_{\pm}^2 = b^2(1-h) + \bar{b}^2(1-\bar{h}) \pm \sqrt{b^4(1-h)^2 + \bar{b}^4(1-\bar{h})^2 + 2 b^2\bar{b}^2 (h\bar{h}+h+\bar{h}-1)}\,.
\eeq
One can easily verify that $G^{-1}$ has a single zero-eigenvalue at $\omega = \pm b$ and $\omega=\pm \bar{b}$. These zero eigenvalues are a consequence of the linearized $SO(2,2)$ gauge symmetry and do not correspond to physical excitations. The other zeros are at
\beq
	\omega^2 - k^2 = m_{\pm}^2\,,
\eeq
and these are genuine propagating excitations. Note that the spectrum of fluctuations is Lorentz-invariant although the action we started with is not manifestly so. Evidently the wormhole is Lorentz-invariant, which perhaps is not so surprising given that the double trace deformation preserves Lorentz symmetry. These are fluctuations of the boundary gravitons, and so we also learn that the deformation gives an effective mass to the boundary gravitons. 

The expressions for the masses-squared $m_{\pm}^2$ are a little complicated. Expressing $b$ and $\bar{b}$ in terms of the double trace coupling $\eta$ leads to an unenlightening expression. Let us focus on the physics. When the operator $O$ is a scalar, with $h=\bar{h} = \frac{\Delta}{2}$, the ensuing wormhole has $b=\bar{b}$ and the masses simplify to
\beq
	m_+^2 = 2b^2\,, \qquad m_-^2 = 2b^2(1-\Delta)\,.
\eeq
Note that the lighter mode becomes tachyonic for $\Delta >1$, when the double trace deformation becomes irrelevant. We conclude that for scalar $O$, the wormhole is stable only for $\Delta <1$. (At $\Delta=1$ the wormhole is singular with $b=\bar{b}=0$.)

\begin{figure}[t]
\centering
\includegraphics[scale=0.6]{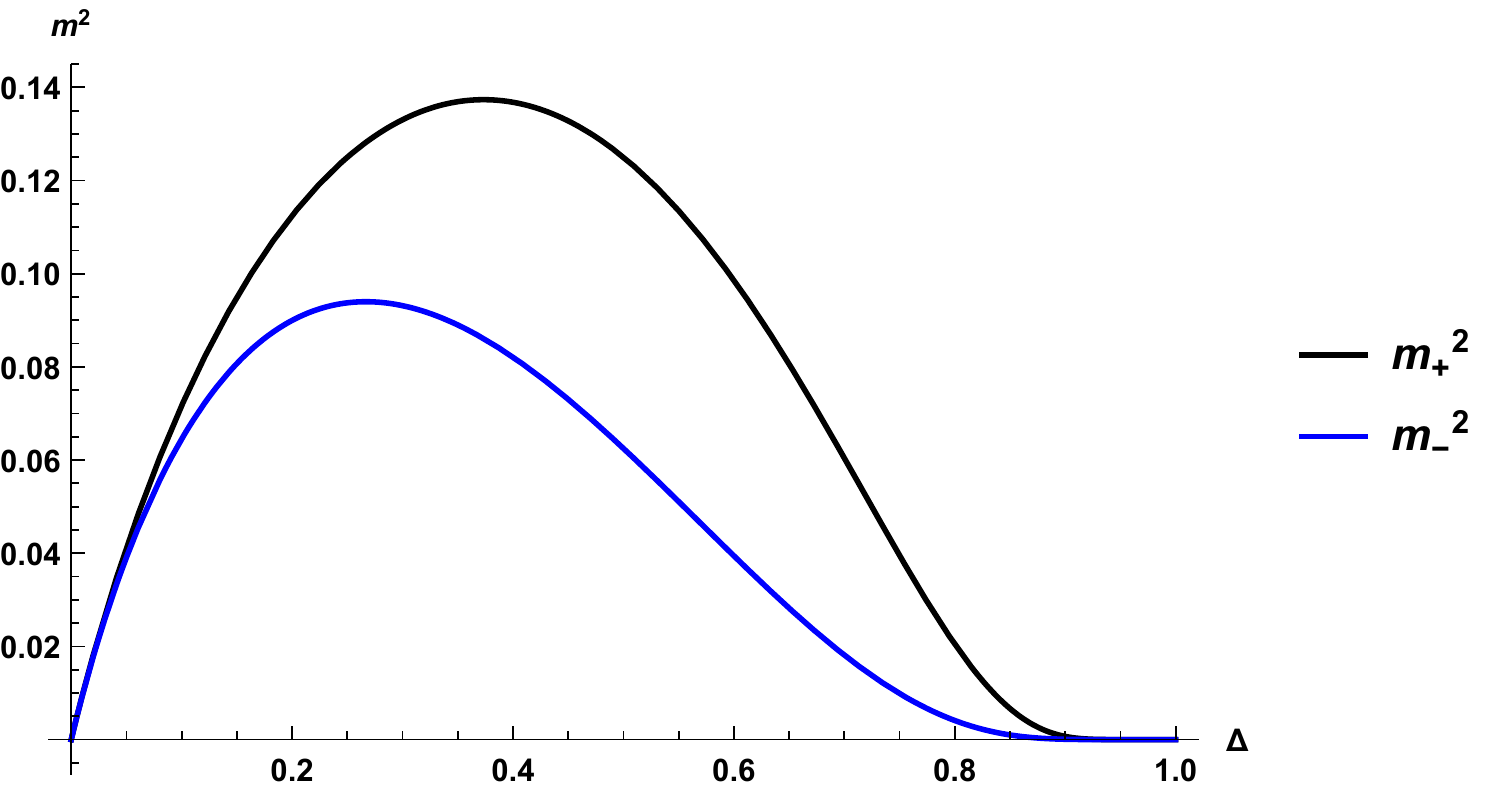}
\caption{\label{F:scalar} The boundary graviton masses when the operator $O$ is a scalar, as a function of $O$'s dimension $\Delta$, for double trace deformation $\eta=1$.}
\end{figure}

\begin{figure}[t]
\centering
\includegraphics[scale=0.6]{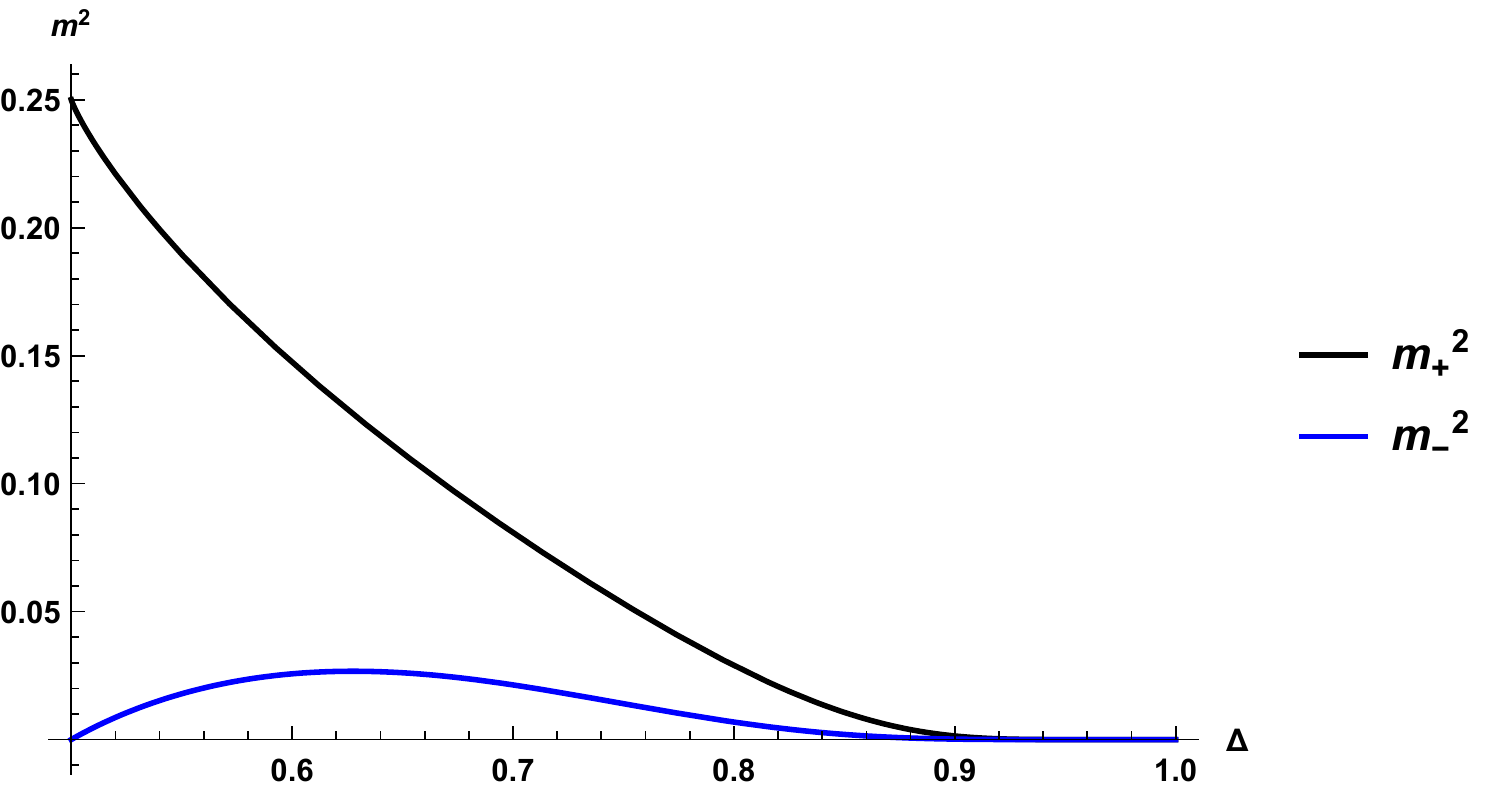}
\caption{\label{F:fermion} The boundary graviton masses when the operator $O$ has spin $s=1/2$, again with $\eta=1$.}
\end{figure}

The same result holds when $O$ carries spin $s = h-\bar{h}$. The lighter mass-squared $m_-^2$ passes through zero precisely when $\Delta=1$, becoming tachyonic for $\Delta >1$. Again the stable wormholes are those stabilized by a relevant double trace. By boundary unitarity $\Delta \geq |s|$ and so the wormhole can only be stabilized by a double trace deformation with spin-0 or spin-1/2 operators. To get a sense of the masses-squared for both cases we plot them in Figs.~\ref{F:scalar} and~\eqref{F:fermion} at fixed value of the double trace parameter $\eta=1$ (the masses-squared are homogenous functions of $\eta$, going as $m^2 \propto \eta^{-\frac{1}{\Delta-1}}$).

\newpage

\subsubsection{$\omega = 0$}

As we mentioned at the beginning of the Section, we have to treat the $\omega=0$ fluctuations separately. The quadratic action for these modes is
\beq
	S_{\rm quad} = \frac{1}{2\pi}\sum_{k\in\mathbb{Z}} \frac{1}{2}\mathbf{\Phi}_0^{\dagger}\cdot G^{-1}_0\mathbf{\Phi}_0\,,
\eeq
where
\beq
\label{E:bdyGravitonPropagator2}
	\mathbf{\Phi}_0= \begin{pmatrix} \delta b(k) \\ \delta \bar{b}(k) \\ \mathcal{Y}(k) \\ \bar{\mathcal{Y}}(k)\end{pmatrix}\,, \qquad G_0^{-1} =\frac{C}{24\pi} \begin{pmatrix} 4(h-1) & 4 \sqrt{h \bar{h}} & -i k & 0 \\ 4 \sqrt{h \bar{h}} & 4(\bar{h}-1) & 0 & i k \\ i k & 0 & \frac{b^2}{2} & 0 \\ 0 & -ik & 0 & \frac{\bar{b}^2}{2}\end{pmatrix}\,.
\eeq
The linearized equations of motion $G_0^{-1}\mathbf{\Phi}_0 = \mathbf{0}$ again have solutions for those $k$ such that $\text{det}\,G_0^{-1}=0$. That determinant is simply
\beq
	\text{det}\,G_0^{-1} =\left( \frac{C}{24\pi}\right)^4 (k^2+m_+^2)(k^2+m_-^2)\,,
\eeq
where $m_{\pm}^2$ were the masses-squared obtained from our analysis at $\omega=0$, given in~\eqref{E:massSquared}.

So there are solutions at
\beq
	\omega^2 - k^2 = m_{\pm}^2\,,
\eeq
for all $\omega$, even $\omega=0$ where the set of fields differs from the set at generic frequency. 

\subsection{Comparing with the disconnected saddle}

In the bulk the boundary conditions we are imposing are that we have two asymptotically AdS$_3$ regions and that there is a negative double trace deformation. The wormhole saddle we have found is not the only one that satisfies these boundary conditions. There is another one, two disconnected copies of global AdS$_3$, and we ought to compare the action of the two saddles to see which is dominant.

The disconnected saddle has a total energy of twice the vacuum energy of 3d gravity, $-\frac{C}{12}$, for a total energy of
\beq
	E_{\rm disconnected} = - \frac{C}{6}\,.
\eeq
Meanwhile the energy of the connected solution is, using the action~\eqref{E:totalS},
\beq
	E_{\rm connected} = \frac{C}{12}\left( b^2 + \bar{b}^2 - \eta b^{2h}\bar{b}^{2\bar{h}}\right) = - \frac{C}{6} \frac{b\bar{b}}{2}\frac{1-\Delta}{\sqrt{h \bar{h}}}\,.
\eeq
where we have used~\eqref{E:useful}. We see that the connected wormhole saddle has lower energy for sufficiently large $b\bar{b}$. Since $b\bar{b}\propto \eta^{\frac{1}{1-\Delta}}$ by~\eqref{E:bSaddle} it follows that the wormhole dominates over disconnected saddle for sufficiently large double trace deformation. There is a phase transition at finite $\eta=\eta_c$ at which $E_{\rm disconnected} = E_{\rm connected}$. For $\eta < \eta_c$ the disconnected saddle dominates.

If we go to finite temperature $T$ there is yet another disconnected saddle, two copies of a BTZ black hole. At zero double trace deformation there is the usual Hawking-Page transition from periodically identified Euclidean global AdS$_3$ to Euclidean BTZ. We expect that transition to persist for $\eta < \eta_c$, while for $\eta>\eta_c$ we expect there to be a first order transition between the wormhole saddle at low temperature and two disconnected copies of BTZ at high temperature.

\subsection{One-loop determinant}

By the by, the simplicity of the determinant of the quadratic action,~\eqref{E:det}, can be used to easily extract the one-loop determinant of boundary gravitons around the wormhole saddle. The appropriate measure for the $\phi_i$ and $\bar{\phi}_j$ is, using the symplectic measure associated with the boundary graviton theory~\cite{Cotler:2018zff,Cotler:2020ugk}, 
\beq
	[d\phi_1 d\phi_2 d\bar{\phi}_1d\bar{\phi}_2] \propto \prod_{k\in \mathbb{Z}} \prod_{\omega\in \mathbb{R}} \frac{\omega^2|\omega^2-b^2| |\omega^2-\bar{b}^2|}{b^2\bar{b}^2} d\phi_1(k,\omega)d\phi_2(k,\omega)d\bar{\phi}_1(k,\omega) d\bar{\phi}_2(k,\omega)\,.
\eeq
Those factors of $\omega^2, |\omega^2-b^2|$, and $|\omega^2-\bar{b}^2|$ cancel those appearing in the one-loop determinant $\sim 1/\sqrt{\text{det }G^{-1}}$ coming from~\eqref{E:det}, so that the effective one-loop determinant is
\beq
	\frac{\omega^2|\omega^2-b^2||\omega^2-\bar{b}^2|}{\sqrt{\text{det }G^{-1}}} \propto \frac{1}{\sqrt{(\omega^2-k^2-m_+^2)(\omega^2-k^2-m_-^2)}}\,,
\eeq
i.e. that of two decoupled scalar fields with masses-squared $m_{\pm}^2$. The one-loop approximation to the wormhole partition function (periodically identifying imaginary time $t=-i y$ with $y\sim y+\beta$ in the $\beta\to\infty$ limit) is then
\beq
	Z_{\rm wormhole} \approx Z_{+}Z_- e^{-\beta E_{\rm connected}}\,,
\eeq
where $Z_{\pm}$ is the one-loop determinant of a two-dimensional scalar with mass-squared $m_{\pm}^2$. 

\section{Bulk geometry and boundary correlations}
\label{S:3d}

Having identified stable wormhole saddles, we would like to understand some of their physics. First, we attempt to reconstruct the geometry of the bulk wormhole. Second, we study the boundary graviton propagator, which is enough to give tree-level correlations between the two boundaries as well as a first step toward a computation of scattering in the wormhole geometry.

\subsection{Bulk reconstruction}

What is the bulk geometry of the wormhole we found in the last Section? Here we present a simple argument that, when the stabilizing double trace is a scalar operator, it is
\beq
\label{E:bulkGeometry}
	ds^2 = d\rho^2 + b^2 \cosh^2(\rho)(-dt^2+dx^2)\,.
\eeq
The two main ingredients we use are the geometric optics approximation to two-point functions of heavier operators across the wormhole, and the evidence from the fluctuation spectrum that the wormhole respects boundary Lorentz invariance.

First consider the matrix of two-point functions of a different operator $\mathcal{O}$ than the $O$ whose double trace stabilizes the wormhole. Those two-point functions are operators in the boundary graviton theory; the left-right two-point function has the same form as $\langle O_L O_R\rangle$,~\eqref{E:doubleTraceL}, just with the scaling weights $(h,\bar{h})$ replaced by those of $\mathcal{O}$. There are similar expressions for the left-left and right-right two-point functions.

Now suppose that $\mathcal{O}$ is sufficiently heavy that its two-point function may be evaluated in the geometric optics approximation, that is, in terms of bulk geodesics connecting the two boundary insertions. Thinking of $\langle \mathcal{O}_L(t_L,x) \mathcal{O}_R(t_R,x)\rangle$ in terms of a bulk geodesic at fixed $x$, we can reconstruct the bulk geometry at fixed $x$, with the result
\beq
	ds^2 = d\rho^2 - b^2 \cosh^2(\rho)dt^2 + (\text{terms involving }dx)\,.
\eeq
Given the boundary Lorentz invariance there is a natural guess for the spacetime metric, namely~\eqref{E:bulkGeometry}. 

This guess resonates with a simple picture of the double trace deformation as a uniform density of Wilson lines (or, if $O$ is a scalar, of geodesics) connecting the two boundaries. Coupling such a density to 3d gravity would lead to a bulk stress tensor whose only nonzero component would be $T_{\rho \rho}$. Solving the remaining Einstein's equations for a connected ansatz of the form
\beq
	ds^2 = d\rho^2 + e^{2A(\rho)}(-dt^2+dx^2)\,,
\eeq
leads to a line element of the form~\eqref{E:bulkGeometry}, $ds^2 = d\rho^2 + b'^2 \cosh^2(\rho)(-dt^2+dx^2)$, for some parameter $b'$~\cite{Cotler:2021cqa,Cotler:2022rud}. That value can then be matched to the ``$b$'' appearing in the boundary graviton theory.

However it is yet unclear if the matter stress tensor supporting the wormhole indeed takes this form where the only nonzero entry is $T_{\rho \rho}$. It would be nice to see if this is the case from a direct computation of the one-loop renormalized stress tensor for a bulk scalar with a negative double trace deformation. 

\subsection{Boundary correlation functions}

Now let us consider simple boundary correlation functions obtained from these traversable wormholes, focusing on operators constructed from the boundary graviton fields. The most basic object we could consider is the boundary graviton propagator. However, given the non-standard field content of the boundary theory, with some fields only depending on the spatial circle $x$, as well as the pattern of gauge symmetries, we elect to study a closely related object, a certain matrix of two-point functions of gauge-invariant operators. These operators exist on both boundaries, and so the matrix of correlations encodes the propagation of boundary gravitons on one boundary to the other.

Our boundary action has $C=3/(2G)\gg 1$ and $1/C$ is a weak coupling. Here we work in the free field limit.

In the absence of a double trace deformation the boundary graviton action~\eqref{E:lorentzianS} is essentially a sum of four Alekseev-Shatashvili models, two for the $\Phi_i$ and two for the $\bar{\Phi}_i$. (Without a double trace deformation, the ``twist fields'' $Y$ and $\bar{Y}$ ensure that $b$ and $\bar{b}$ are constants, and on such a constant the total action decouples into a sum of four parts each involving a $\Phi_i$ or $\bar{\Phi}_i$.) Each of those models is chiral with a chiral stress tensor, taking the form~\cite{Cotler:2018zff}
\beq
	T_i = \frac{C}{12}\left\{ \tan\left( \frac{\Phi_i}{2}\right),t\right\}\,, \qquad \bar{T}_i = \frac{C}{12}\left\{ \tan\left( \frac{\bar{\Phi}_i}{2}\right),t\right\}\,,
\eeq
where $\{f(x),x\} = \frac{f'''(x)}{f'(x)} - \frac{3}{2}\frac{f''(x)^2}{f'(x)^2}$ is the Schwarzian derivative of $f(x)$ with respect to $x$. These operators are invariant under $x$-dependent $SO(2,2)$ gauge transformations as the Schwarzian derivative is invariant under fractional linear transformations, $\left\{ \frac{a f(t)+b}{cf(t)+d},t\right\} = \left\{ f(t),t\right\}$. With the double trace deformation there are additional contributions to the boundary stress tensor, but for illustrative purposes we consider the $T_i$ and $\bar{T}_j$ as they are the simplest gauge-invariant operators in the theory.

Here we compute the matrix of two-point functions of the $T_i$ and $\bar{T}_j$ to leading order at large $C$, when the double trace deformation involves a scalar operator $O$. We focus on this two-point function at $k=0$ in the time domain.

To linear order in fluctuations around the wormhole saddles considered in this work, we have
\begin{align}
\begin{split}
	T_i & = \frac{Cb^2}{24} +\frac{C}{12}\left( b \delta b +\left( \frac{\partial_t^3 \phi_i}{b} + b \partial_t \phi_i\right)\right)  + O(\delta\Phi^2)\,,
	\\
	\bar{T}_i & = \frac{C\bar{b}^2}{24}+ \frac{C}{12}\left( \bar{b}\delta \bar{b} + \left( \frac{\partial_t^3 \bar{\phi}_i}{\bar{b}} + \bar{b} \partial_t \bar{\phi}_i\right)\right) + O(\delta\bar{\Phi}^2)\,.
\end{split}
\end{align}
In the Fourier domain and using that $\bar{b}=b$ for scalar $O$ the linearized fluctuations of $T_i$ and $\bar{T}_i$ are
\begin{align}
\begin{split}
	\delta T_i(k,\omega) & = \frac{C}{12} \left( 2\pi \delta(\omega) \,b \delta b(k) + \frac{i\omega}{b}(\omega^2-b^2)\phi_i(k,\omega)\right)\,,
	\\
	\delta \bar{T}_i(k,\omega) & = \frac{C}{12}\left( 2\pi \delta(\omega) \,b \delta \bar{b}(k) + \frac{i\omega}{b}(\omega^2-b^2)\bar{\phi}_i(k,\omega)\right)\,.
\end{split}
\end{align}
At nonzero frequency we see that the connected two-point function of the $T_i$ and $\bar{T}_j$ is proportional to that of the $\phi_i$ and $\bar{\phi}_j$, e.g.
\beq
	\langle T_i(-k,-\omega) T_j(k,\omega)\rangle = \left( \frac{C}{12}\right)^2\left(  \omega^2(\omega^2-b^2)^2 \langle \phi_i(-k,-\omega)\phi_j(k,\omega)\rangle + O\left(\frac{1}{C}\right)\right)\,,
\eeq
while at zero frequency e.g.
\beq
	\langle T_i(-k)T_j(k)\rangle = \left( \frac{C}{12}\right)^2 \left( b^2 \langle \delta b(-k)\delta b(k)\rangle + O\left( \frac{1}{C}\right)\right)\,.
\eeq
(All four of these correlation functions coincide to leading order in large $C$, which is why there is no index structure on the RHS.)

Let us package the $T_i$ and $\bar{T}_j$ into a four-component vector $\mathcal{T}_a= (T_1, T_2 , \bar{T}_1, \bar{T}_2)$, and consider the Fourier-space two-point function
\beq
	\mathcal{G}_{ab}(k,\omega) =\langle \mathcal{T}_a(-k,\omega)\mathcal{T}_b(k,\omega)\rangle \,.
\eeq
To obtain this at nonzero frequency and zero spatial momentum we use the propagator for the $\phi_i$ and $\bar{\phi}_j$ which is given by the matrix $iG$ in~\eqref{E:bdyGravitonPropagator}, while at zero frequency we use the propagator for $\delta b$ and $\delta \bar{b}$ as obtained from the matrix $iG_0$ in~\eqref{E:bdyGravitonPropagator2}. After some simplification we find
\begin{align}
\label{E:TTfrequency}
	\mathcal{G}_{ab}(k=0,\omega) &= \frac{\pi C b^2}{12}\frac{i}{(-\omega^2+m_+^2)(-\omega^2+m_-^2)}
	\\
	\nonumber
	& \qquad \times \left( -2h(\omega^2-b^2)^2\Delta_{ab} + (\omega^2-m_-^2)\begin{pmatrix} X(\omega) & b^2 & 0 & 0 \\ b^2& X(\omega) & 0 & 0 \\ 0 & 0 & X(\omega) & b^2 \\ 0 & 0 & b^2 & X(\omega) \end{pmatrix}+O\left( \frac{1}{C}\right)\right)\,,
\end{align}
where $\Delta_{ab}$ is the matrix whose entries are $1$ for all $a$ and $b$, and
\beq
	X(\omega) = \frac{2\omega^2(\omega^2-m_+^2)}{b^2}+b^2\,.
\eeq
Note that there are poles only at the physical masses-squared $m_{\pm}^2$. 

The zero-frequency limit of this object agrees with the two-point function at zero frequency, obtained instead from the propagator for $\delta b$ and $\delta\bar{b}$, e.g.
\beq
	\mathcal{G}_{11}(k=0,\omega=0) \approx i \left(\frac{C}{12}\right)^2 b^2 \langle \delta b(k=0)\delta b(k=0)\rangle \approx - \frac{\pi i Cb^2}{24} \frac{1-h}{1-2h}\,.
\eeq
That is the expression~\eqref{E:TTfrequency} holds at all $\omega$.

Inverse Fourier transforming back to real time,\footnote{It is helpful to think of this object as the spatially integrated two-point function of the $T_i$ and $\bar{T}_j$, with $\mathcal{G}_{ab}(t) = \int_0^{2\pi} dx\, \mathcal{G}_{ab}(t,x)$.}
\beq
	\mathcal{G}_{ab}(t) = \int \frac{d\omega}{2\pi } e^{-i\omega t} \mathcal{G}_{ab}(k=0,\omega)\,,
\eeq
the time-ordered two-point function is
\beq
	\mathcal{G}_{ab}(t) = \frac{ \pi C b^4}{48}\text{sgn}(t)\left( \frac{(1-4h)^2e^{-im_-|t|}}{m_-}\Delta_{ab}+\frac{e^{-im_+|t|}}{m_+}\begin{pmatrix} 1 & 1 & -1 & -1 \\ 1 & 1 & -1 & -1 \\ -1 & -1 & 1 & 1 \\ -1 & -1 & 1 & 1  \end{pmatrix} + O\left( \frac{1}{C}\right)\right)\,.
\eeq
This matrix has off-diagonal components indicating correlations between fields on the different boundaries. Note also that the commutator of operators on the different boundaries does not vanish, as expected, indicating that these correlations are due to interactions in the Hamiltonian rather than to correlations in the particular state at hand. 

\section{Discussion}
\label{S:discuss}

In this work we have found traversable wormholes in three-dimensional gravity. These wormholes connect two asymptotically AdS$_3$ regions and are stabilized by an eternal version of the Gao-Jafferis-Wall deformation. Alternately they are a three-dimensional version of the traversable wormholes in JT gravity found by Maldacena and Qi~\cite{Maldacena:2018lmt}.

Our analysis relied on two main ingredients. The first was the reduction of three-dimensional gravity with negative cosmological constant on spacetimes with the topology of interest (an annulus times time) to a boundary theory. That boundary model is the continuation of that obtained in~\cite{Cotler:2020ugk} in the context of Euclidean wormholes. Roughly speaking, it is four copies of the Alekseev-Shatashvili model~\cite{Alekseev:1988ce,Cotler:2018zff}, two for each boundary, giving an effective field theory description of the boundary stress tensor. The second was the two-point function of light operators, with one insertion on each boundary, expressed as an operator in the boundary graviton model. That operator is a bilocal operator in the Alekseev-Shatashvili-like model, conceptually similar to the bilocal operators of the Schwarzian model. With this operator in hand, we obtained a boundary action for the traversable wormhole by putting these two ingredients together. Thanks to the negative double trace deformation we found a stable saddle, at least when the double trace deformation was relevant. By studying fluctuations around the wormhole saddle we saw that the wormhole was in fact invariant under boundary Lorentz transformations and that boundary gravitons gain a mass thanks to the deformation.

The bulk model we require, gravity coupled to a large number of scalars dual to relevant operators, is rather contrived and unrealistic. While this work has little to do with a generic AdS$_3$ compactification of string theory we note that the supersymmetric 2d SYK model of Murugan, Stanford, and Witten~\cite{Murugan:2017eto} offers an arena with a large $c$ CFT and a large number of light operators where one can put these ideas to work. To the extent that model has a gravity dual, it is ``stringy,'' with light higher spin states, but even so we expect that there is a ``traversable wormhole'' saddle in two copies of that model, like that in  the standard SYK model studied by Maldacena and Qi.

Because we work entirely on the boundary our analysis is unable to identify the precise geometry of the bulk wormhole. By computing boundary correlation functions we can see that the two AdS$_3$ boundaries are causally connected, and while we gave a plausible candidate for the bulk geometry thanks to those boundary correlators, it would be nice to have a direct derivation using the one-loop stress tensor of bulk matter fields with an eternal Gao-Jafferis-Wall deformation.

Relatedly, in the context of a momentary rather than eternal inter-boundary coupling, Gao and Liu~\cite{Gao:2018yzk} have shown how to compute correlation functions encoding propagation and regenesis through the ensuing wormhole directly from considerations in large $c$ CFT, resumming conformal perturbation theory in the negative double trace deformation. Can those methods be generalized to describe an eternal negative double trace coupling, making contact with the wormholes studied in this work?

\subsection*{Acknowledgements}

We would like to thank J.~Cotler and M.~van Raamsdonk for useful discussions. WH and KJ are supported in part by an NSERC Discovery Grant.

\bibliography{refs}
\bibliographystyle{JHEP}

\end{document}